\documentclass[runningheads]{llncs}

\usepackage{booktabs}
\usepackage{colortbl}
\usepackage[T1]{fontenc}
\usepackage{graphicx}
\usepackage{lipsum}
\usepackage{paralist}
\usepackage{xcolor}

\newcommand\blfootnote[1]{%
  \begingroup
  \renewcommand\thefootnote{}\footnote{#1}%
  \addtocounter{footnote}{-1}%
  \endgroup
}

\begin{document}

\title{Enabling Seamless Transitions from Experimental to Production HPC for Interactive Workflows}

\titlerunning{Interactive HPC Workflows from Testbed to Production}

\author{
    Brian D. Etz\orcidID{0000-0002-0855-4863} \and
    David M. Rogers\orcidID{0000-0002-5187-1768} \and
    Michael J. Brim\orcidID{0000-0002-7479-5526} \and
    Ketan Maheshwari\orcidID{0000-0001-6852-5653} \and
    Kellen Leland\orcidID{0000-0001-7460-3064} \and
    Tyler J. Skluzacek\orcidID{0000-0003-2242-4931} \and
    Jack Lange\orcidID{0000-0003-0616-7437} \and
    Daniel Pelfrey\orcidID{0000-0002-5625-1373} \and
    Jordan Webb\orcidID{0000-0003-3887-7601} \and
    Patrick Widener\orcidID{0000-0002-5882-0816}\and
    Ryan Adamson\orcidID{0000-0002-7194-9577} \and
    Christopher Zimmer\orcidID{0000-0001-5054-4354} \and
    Verónica G. Melesse Vergara\orcidID{0000-0002-4333-4145} \and
    Mallikarjun Shankar\orcidID{0000-0001-5289-7460} \and
    Sarp Oral\orcidID{0000-0001-8745-7078} \and
    Rafael Ferreira da Silva\orcidID{0000-0002-1720-0928}
}

\authorrunning{B. Etz et al.}

\institute{Oak Ridge National Laboratory, Oak Ridge, TN 37831, USA}

\maketitle

\blfootnote{This manuscript has been authored in part by
    UT-Battelle, LLC, under contract DE-AC05-00OR22725 with the US
    Department of Energy (DOE). The US government retains and the
    publisher, by accepting the article for publication, acknowledges
    that the US government retains a nonexclusive, paid-up,
    irrevocable, worldwide license to publish or reproduce the
    published form of this manuscript, or allow others to do so, for
    US government purposes. DOE will provide public access to these
    results of federally sponsored research in accordance with the DOE
    Public Access Plan
    (\url{http://energy.gov/downloads/doepublic-access-plan}).}

\begin{abstract}
The evolving landscape of scientific computing requires seamless transitions from experimental to production HPC environments for interactive workflows. This paper presents a structured transition pathway developed at OLCF that bridges the gap between development testbeds and production systems. We address both technological and policy challenges, introducing frameworks for data streaming architectures, secure service interfaces, and adaptive resource scheduling for time-sensitive workloads and improved HPC interactivity. Our approach transforms traditional batch-oriented HPC into a more dynamic ecosystem capable of supporting modern scientific workflows that require near real-time data analysis, experimental steering, and cross-facility integration.

\keywords{Interactive HPC Workflows \and Time-Sensitive Computing \and Cross-Facility Integration \and High Performance Computing.}
\end{abstract}

\section{Introduction}

High performance computing (HPC) has traditionally operated in a batch-oriented paradigm, where jobs are submitted to queues, executed when resources become available, and results are retrieved after completion~\cite{10.1007/978-3-031-73716-9_16}. However, modern scientific workflows increasingly demand interactive capabilities that allow researchers to steer computations in near real time, analyze data as it is generated, and make informed decisions during the execution of experiments~\cite{ferreiradasilva2024computer,reuther2024interactiveurgenthpcchallenges}. This shift is driven by scientific machine learning (ML) models that guide exploration and by experimental facilities that produce data at unprecedented rates~\cite{jha2022ai,antypas2021enabling}. Therefore, as these workflows mature, alternative interactive computing strategies that complement traditional computing operations are required for seamless integration across facility boundaries.

Time-sensitive applications present particularly challenging requirements for HPC systems. These applications require time-critical/sensitive access to computing resources (i.e., real time or near real time), motivated by a need for timely decision making, experiment steering, virtual proximity, or loss of data fidelity~\cite{IRI-ABA-Report,reuther2024interactiveurgenthpcchallenges}. Experiments such as the Linac Coherent Light Source (LCLS-II) require near real-time data analysis and instrument steering within strict time constraints~\cite{10.1051/epjconf/202429513002}. This requires preemptive scheduling and advance reservations to align computational resources with experiment timelines. Similarly, data-intensive integration workflows for DIII-D tokamak fusion experiments~\cite{10.3389/fphy.2024.1524041} require robust data movement and flexible authentication across institutional boundaries. Current HPC technologies and policies, designed primarily for batch processing, are often ill-equipped to support these emerging patterns of scientific computing.

Testbeds offer an attractive approach to evaluate and address interactive HPC challenges, yet a significant barrier remains between testbed and production environments. While testbeds provide flexibility for innovative workflows, they lack the scale, reliability, and security required for production science. Conversely, production HPC systems are optimized for traditional batch workloads with policies that often impede interactive applications. This disconnect forces researchers to substantially modify workflows when transitioning to production from sandboxed testbed environments, slowing scientific progress and limiting the adoption of cutting-edge methodologies.

\begin{figure}[!t]
    \centering
    \includegraphics[width=\linewidth]{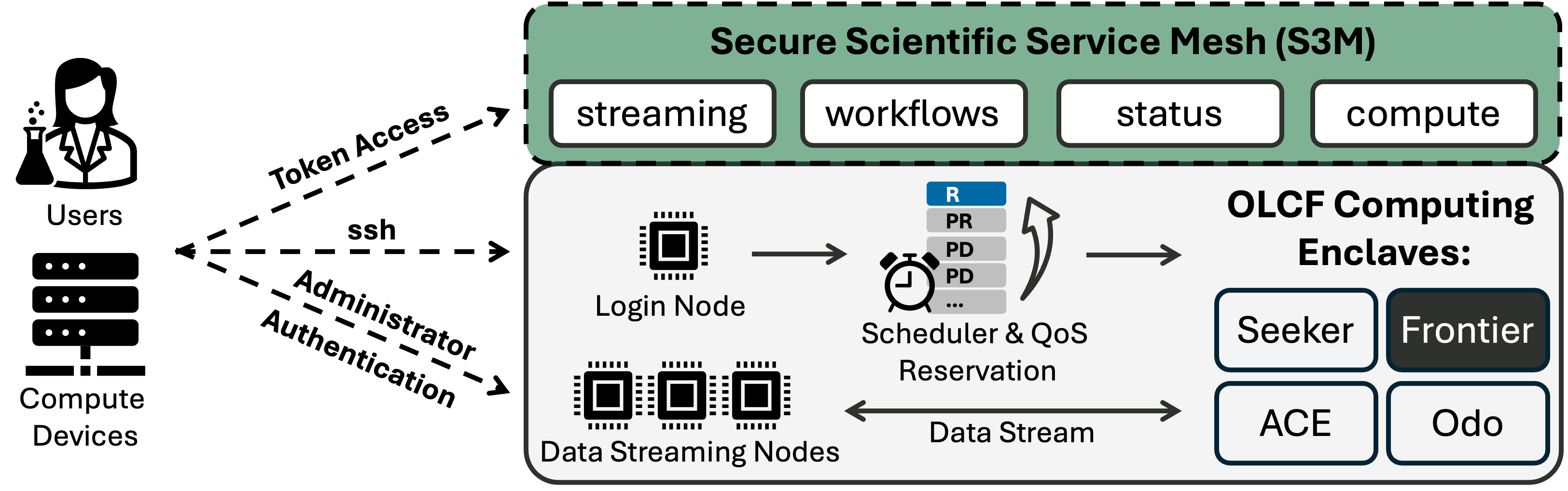}
    \caption{OLCF's integrated approach to interactive computing. External users connect via secure authentication methods, accessing a suite of specialized services for status monitoring, computation, workflow management, and data streaming that collectively enable seamless interactivity across the computing environments.}
    \label{fig:approach}
\end{figure}

At the Oak Ridge Leadership Computing Facility (OLCF), we are establishing a transition pathway that bridges the gap between development testbeds and production HPC environments (Section \ref{sec:transition}). This pathway provides a rubric for researchers to easily and systematically adapt their interactive workflows through progressively enhanced computing environments, while at the same time providing useful information on how OLCF might need to adapt our production environments in the future. Our approach is motivated by close collaboration with users to enable their interactive workflows and combines the necessary policy adaptations for security, operations, and resource allocation with important technological innovations in data streaming, secure service interfaces, and flexible resource scheduling (Fig.~\ref{fig:approach}). This paper examines these policy and technological requirements, using a practical case study to demonstrate effective strategies for seamless transitions from experimental to production environments.

\section{End-to-End Interactive Workflow}
\label{sec:example}

Modern scientific computing demands both specialized infrastructure and efficient data exchange mechanisms. This section presents OLCF's computing environments for interactive workflow development and introduces a project that implements experimental data streaming for near real-time analysis.

\subsection{Computing Landscape at OLCF}
\label{sec:OLCF}

OLCF provides a spectrum of computing environments designed to facilitate transitions from experimental to production workflows. Frontier~\cite{frontier}, the world's first exascale system, delivers unprecedented performance while operating within traditional batch-oriented HPC paradigms. Frontier serves as the ultimate destination for mature workflows that require leadership-class computing capabilities (i.e., jobs consuming 20\% or more compute nodes at any given time).

To bridge the development-to-production gap, OLCF has established the Advanced Computing Ecosystem (ACE)~\cite{aceiri24} and Quantum Computing User Program (QCUP).These environments enable researchers to explore interactive workflows across facilities, test emerging technologies, and evaluate necessary policy adaptations before transitioning to production systems. ACE enables significant progress in technology development and supports diverse computing patterns identified by DOE's Integrated Research Infrastructure (IRI) program~\cite{IRI-ABA-Report}, while QCUP allows exploration of quantum computing integration with traditional HPC resources~\cite{10.1016/j.future.2024.06.058}.

ACE has played a pivotal role in advancing interactive HPC capabilities by providing a flexible environment for experimentation with emerging technologies and collaboration with external users across experimental and computational facilities. The testbed supports diverse computational patterns, allowing researchers to prototype innovative workflows that require near real-time data analysis, experimental steering, and cross-facility integration. ACE's infrastructure accommodates the development of custom authentication mechanisms, data streaming architectures, and specialized resource scheduling required by interactive applications. This environment has proven particularly valuable for time-sensitive workloads like the LCLS-II experiments, where near real-time analysis of experimental data guides ongoing scientific discoveries. By creating a sandbox environment, 
ACE enables rapid iteration on complex multi-facility workflows before they must contend with the stricter policies of production systems.

\subsection{Streaming Experimental Data for Near Real-time Analysis}
\label{sec:lclstream}

The LCLStream project exemplifies the challenges of transitioning interactive workflows to production environments, aiming to accelerate scientific discovery by processing beamline data from SLAC's LCLS facility on OLCF's HPC resources~\cite{10.3389/fhpcp.2025.1536471}. This workflow enables researchers to launch compute jobs during their scheduled beamtime or interactive analysis sessions, establishing high-bandwidth data paths between the data transfer nodes and allocated HPC resources. 

Despite its conceptual simplicity, the LCLS implementation encountered three fundamental challenges that would later inform our holistic framework for interactive HPC:

\begin{compactenum}
    \item High-performance data movement required innovations in network architecture. For LCLStream's high-speed data transfer needs, we implemented experimental network configurations that connected the Energy Sciences Network (ESNet) testbed and science DMZ networks to ACE compute nodes, with carefully managed firewall rules and SLURM integration to maintain security boundaries. Although successful in achieving full switching capacity throughput, these approaches proved too complex for production environments, prompting the development of the memory-to-memory data streaming framework described in Section~\ref{sec:mem-to-mem} as a more scalable solution.

    \item Authentication and access control emerged as a critical barrier. OLCF security models traditionally rely on human-verified SSH sessions rather than programmatic API access, creating a significant obstacle for automated workflows that require machine-to-machine communication across facility boundaries. To address this challenge, we implemented a container-based solution to deploy custom REST APIs that satisfy security requirements while enabling programmatic resource access. These APIs allow LCLS users to authenticate once through multifactor verification and delegate secure access tokens to their software, a pattern that directly informed the development of our Secure Scientific Service Mesh (S3M) framework detailed in Section~\ref{sec:s3m}.

    \item Time-sensitive resource allocation required fundamental rethinking of scheduling models. In an environment optimized for large batch jobs, interactive workflows with strict timing requirements demanded innovative approaches to resource prioritization. Our solution implemented on ACE utilizes SLURM's preemptive scheduling capabilities, refined through a ``Quality of Service (QoS) reservation" system that provides guaranteed availability without the resource inefficiency of traditional node reservations. This approach, detailed in Section~\ref{sec:time-sensitive}, allows specific user groups to receive elevated priority during scheduled periods, ensuring that their workflows start immediately while minimizing system-wide impact.
\end{compactenum}



\medskip

The LCLStream experience directly shaped our comprehensive approach to interactive HPC workflows. Through this project, we discovered that successful cross-facility integration requires simultaneous evolution in three interconnected domains: security, operations, and resource allocation. On the security front, we had to reimagine authentication from a user-centric model to a service-oriented framework with delegated credentials and fine-grained access controls, preserving strong security boundaries while enabling automated data flows. Operationally, the project revealed a fundamental tension between the unpredictable timing of experimental facilities and the rigid maintenance schedules of production HPC systems, necessitating new service level agreements and incident response protocols specifically designed for time-sensitive workloads. Perhaps most significantly, LCLStream challenged traditional resource allocation metrics based solely on system utilization, highlighting the scientific value of immediate but short-duration access to computing resources aligned with experimental beamtime. These insights, together with similar requirements from other applications, formed the foundation of our policy considerations (Section~\ref{sec:policies}) and drove the development of technologies that could systematically address these interrelated challenges while providing a clear pathway from experimental to production deployment.

As these solutions mature, they establish patterns to address similar requirements in other interactive HPC applications transitioning to production environments, and guide our structured transition pathway detailed in Section~\ref{sec:transition}. The LCLStream project demonstrates how each stage of the transition from ACE to production systems requires careful consideration of both technological capabilities and policy frameworks to maintain workflow functionality while enhancing security and reliability.

\section{Policy Considerations and Challenges}
\label{sec:policies}

Enabling interactive HPC workflows requires fundamental policy adaptations beyond the technological solutions. Traditional HPC policies, designed primarily for batch processing, must evolve to accommodate the dynamic and time-sensitive nature of interactive workflows while ensuring system security and efficiency.

\subsection{Security Policy Adaptations}
\label{sec:policy-security}

Security policies are designed to ensure high assurance for the appropriateness of data access and correctness of workflow execution. Interactive workflows challenge traditional HPC security models by requiring external data streams to access supercomputing resources on demand. This represents a significant departure from conventional isolated computing environments, where access is mediated exclusively through SSH sessions. For projects like LCLStream, where experimental facilities need to initiate workflows and stream data to computing resources in real-time, security policies must be reimagined while ensuring every user and their actions are authenticated, authorized, and evaluated for appropriate use and intent for the corresponding computing environment.

We have developed a security framework that implements token-based authentication with time-limited credentials alongside API gateways on secure access nodes with fine-grained authorization controls. Application-aware traffic governance through S3M (Section~\ref{sec:s3m}) provides an additional layer of protection. This approach maintains robust security boundaries while enabling the necessary external connections for interactive science.

Our experience indicates that traditional network firewalls often become bottlenecks for interactive workloads. Instead, science DMZ architectures with dedicated network interfaces for interactive services provide better performance while maintaining security through isolated VLANs and strict access controls. Security protocols must evolve from a closed fortress model to a selectively permeable membrane that allows data and computations to flow as needed for science, but under tight scrutiny and control.

\subsection{Operational Policy for Time-Sensitive Workloads}

Interactive workflows require operational guarantees that conflict with traditional maintenance and resource allocation policies. To accommodate these needs, maintenance windows must be carefully scheduled, communicated, and tightly coupled with resource reservations, with provisions for critical interactive services to remain operational, when possible. Service Level Agreements (SLAs) for interactive services with defined uptime requirements provide clear expectations for both users and system administrators.

Specialized incident response protocols now prioritize restoration of time-sensitive services when issues arise. Change management procedures include an additional risk assessment for modifications that could affect interactive services, with increased testing requirements and fallback provisions. These operational adaptations recognize that interruptions to interactive workflows can have major scientific consequences, especially when coordinated with experimental facilities.




\subsection{Resource Allocation and Scheduling Policies}

Perhaps the most challenging policy shift involves fairly allocating resources between traditional batch jobs and interactive workflows with strict timing requirements. Our approach includes QoS-based preemptive scheduling tiers that enable time-critical applications to interrupt lower-priority workloads when necessary. Resource allocation and scheduling mechanisms guarantee availability without the inefficiency of traditional node reservations by elevating priority for specific user groups during scheduled periods.

Multi-tenancy policies now allow appropriate resource sharing while maintaining security and performance boundaries between users. In experimental environments, containerization technologies like Docker and OpenShift provide sufficient isolation, but production HPC systems require more robust mechanisms such as virtual machines and confidential computing enclaves to guarantee both security and performance isolation. These differences require policy frameworks that can be translated across enforcement mechanisms as workflows transition from testbed to production. We have implemented formalized justification processes for requesting prioritized access to balance system-wide resource utilization with the needs of time-sensitive applications, with well-defined authorization workflows for accessing shared resources.

The transition from experimental to production environments involves progressively stricter resource allocation policies. While testbeds such as ACE allow flexible, on-demand access, production systems require structured approval processes that balance the needs of interactive workflows against broader system utilization. Resource allocation committees must develop new evaluation criteria that account for the value of time sensitivity, rather than focusing exclusively on core hours or node utilization metrics that have traditionally governed HPC allocation decisions.

\section{Technology Framework for Interactive HPC}
\label{sec:framework}

After addressing policy considerations in Section~\ref{sec:policies}, we now present the technological frameworks required to implement these policies effectively. Our framework balances innovation with security, reliability, and the scale required for production operations. This section describes three core components that collectively enable interactive workflows to successfully transition from testbeds to production environments (Fig.~\ref{fig:approach}).

\subsection{Memory-to-Memory Data Streaming Framework}
\label{sec:mem-to-mem}

Interactive HPC applications increasingly require near real-time data movement between experimental facilities and computational resources as outlined for LCLStream (Section \ref{sec:lclstream}). Our data streaming architecture~\cite{streaming} enables direct memory-to-memory transfers between producer and consumer endpoints, significantly reducing latency compared to traditional file-based transfers, while providing capabilities to connect compute jobs with experimental control applications. This architecture is accessed through an API embedded in S3M (Section 4.2) that provisions RabbitMQ or Redis messaging services on dedicated data streaming nodes (DSNs).



 A defining feature is separation of the control and data planes to maintain security while maximizing performance. Control interfaces allow system administrators to implement authentication mechanisms and network access policies independently of high-throughput data paths. DSNs handle these data paths, operating either as application gateways with application-level context awareness (OSI layer 7) or as dedicated routers with selective traffic forwarding (OSI layer 4). The application gateway approach provides better security control but sacrifices some performance, while the router configuration offers higher throughput but relies on firewall rules rather than the application context.


DSNs connect to both external and internal HPC networks through high-speed interfaces, creating a controlled pathway between these environments. System administrators manage DSN configurations through templates, allowing application teams to specify limited parameters such as allowed external addresses, targeted internal nodes, and buffer settings. This maintains control and security while offering the necessary flexibility for diverse interactive workloads.

\subsection{Cross-Facility Authentication and API Gateway}
\label{sec:s3m}

The Secure Scientific Service Mesh (S3M)~\cite{aceiri24} builds on existing scientific APIs (e.g. Superfacility ~\cite{Enders:2020qbo} and FirecREST~\cite{9308083}) but distinguishes itself through a service mesh architecture that provides a unified framework for customizable services not possible in traditional API gateways such as, policy enforcement, dynamic routing, authentication, authorization, and traffic management between external services and internal HPC resources. A user’s interaction with S3M is generally as follows: (1) access token generation through my OLCF web platform, (2) embed token in requests from client applications, and (3) S3M verifies the user and project permissions, authorizes, and executes the requested task.


This approach enables fine-grained access control to HPC resources through standardized APIs for low-latency data streaming and seamless workflow orchestration, allowing external workflows to interact with computation and storage resources without compromising security boundaries. S3M incorporates policy-as-code capabilities for defining secure access patterns, implements rate limiting to prevent resource exhaustion, and provides comprehensive logging for security auditing. The mesh architecture abstracts the underlying complexity of cross-facility authentication, allowing scientific workflows to focus on core functionality while security policies are consistently enforced at the infrastructure level.



\subsection{Time-Sensitive Workload Prioritization System}
\label{sec:time-sensitive}

Supporting time-sensitive interactive workflows requires fundamental changes in resource scheduling approaches. Our adaptive scheduling framework incorporates preemptive capabilities that allow high-priority interactive jobs to interrupt lower-priority batch workloads when necessary. This ensures resource availability for time-critical analysis while maintaining overall system utilization.

QoS implementations define service levels with associated resource guarantees, allowing administrators to assign appropriate priorities to different workflow types. SLURM's QoS features have been tailored to assign varying levels of priority and preemption capabilities to specific projects based on their time-sensitivity requirements. Users can select the appropriate QoS level when submitting jobs, with a project-based inheritance of these capabilities.

For workflows with more predictable schedules, we have implemented a refined ``QoS reservation" system that provides guaranteed availability without the resource inefficiency of traditional node reservations. This approach requires coordination between users and system administrators to elevate specific user groups priority during scheduled periods, ensuring that their workflows start immediately while minimizing system-wide impact. Our validation test of the SLURM workflow manager scheme demonstrated that this approach effectively balances interactive and batch workloads across diverse user communities~\cite{aceiri24}.

\section{Experiment to Production Transition Architecture}
\label{sec:transition}


We have devised a practical transition guide that researchers can follow to navigate the path from early prototyping to full-scale production deployment. By understanding the key considerations at each stage of this journey, developers can design their interactive workflows to minimize rework and maximize scientific productivity across the complete computing spectrum. This pathway can be conceptualized as a progression through distinct enclaves: open development environments, open production systems, and moderate security production platforms, each with specific requirements and capabilities.

\begin{figure}[!ht]
    \centering
    \vspace{-15pt}
    \includegraphics[width=.9\linewidth]{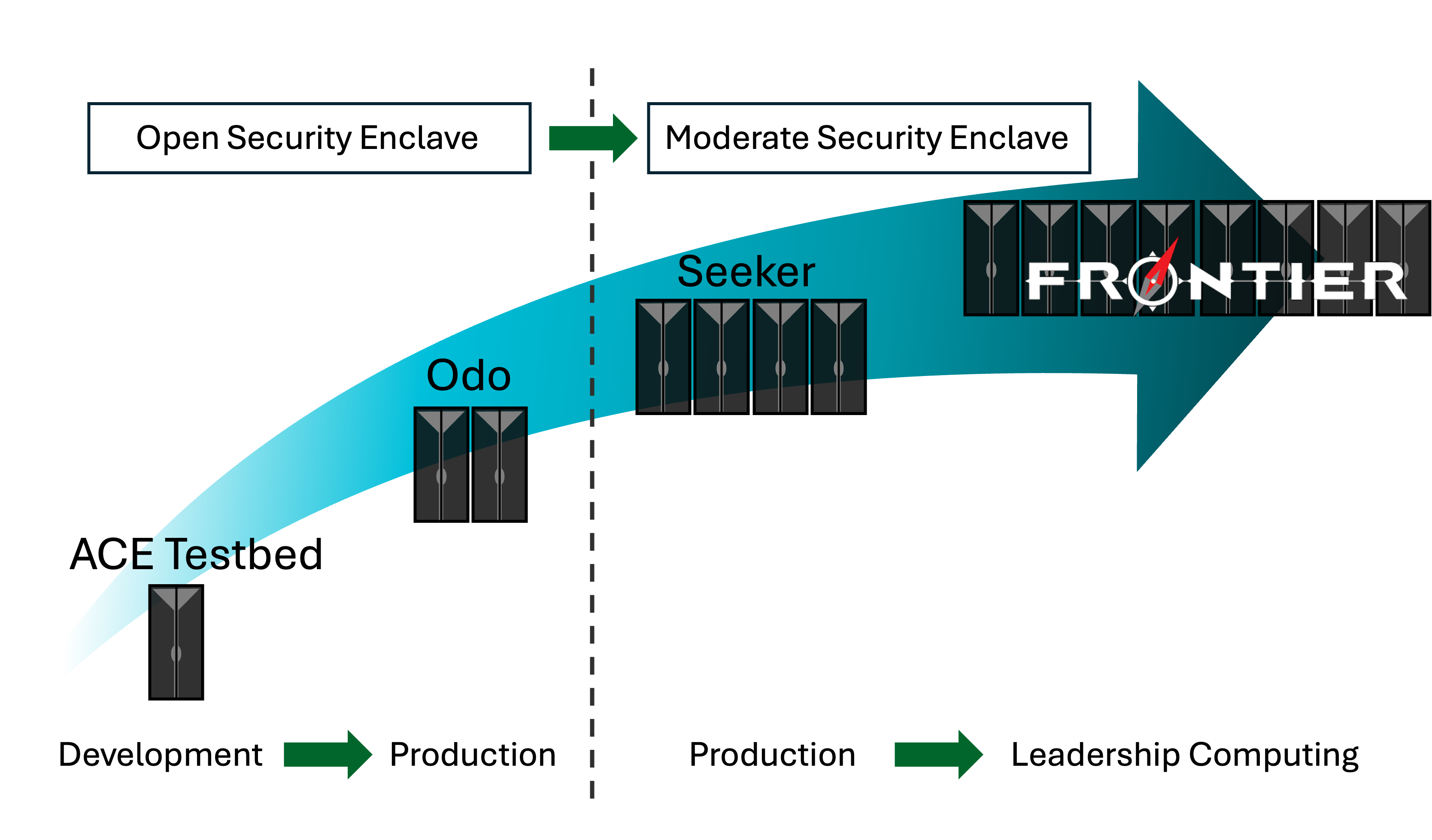}
    \caption{OLCF transition path illustrating progression from the developmental testbed to production HPC to leadership computing, while spanning  security enclaves.}
    \label{fig:GradPlan}
\end{figure}

Open development environments provide flexibility for rapid prototyping and experimentation, allowing researchers to establish workflow patterns and verify technical approaches. As workflows mature, they transition to open production systems, where operational stability becomes more critical, although authentication requirements remain relatively lightweight. The final stage involves deployment to moderate security production environments, where multi-factor authentication and stricter security policies are enforced to protect high-value computational resources. At OLCF, this pathway is implemented through a series of platforms: ACE provides the experimental testbed environment, Odo offers a production-like development system in the open enclave, Seeker serves as a production-development system in the moderate security enclave, and Frontier represents the ultimate production environment (Fig.~\ref{fig:GradPlan}). 

This transition path provides a rubric to production; however, each use case is not built the same and may require different technology and policy considerations. Therefore, each step must be tailored to the specific scope of the project and requires careful consideration of authentication mechanisms, API access patterns, and resource allocation strategies to ensure that workflows remain functional while meeting increasing security requirements. This structured progression enables researchers to gradually evolve their workflows to adapt to the constraints of each environment without the need to re-engineer their workflows.

\medskip

\begin{table}[!ht]
    \scriptsize
    \centering
    \label{tab:summary}
    \caption{Interactive HPC Technologies and Associated Policy Considerations}
    \begin{tabular}{|p{0.15\textwidth}|p{0.62\textwidth}|p{0.21\textwidth}|}
    \hline
    \textbf{Technology} & \textbf{Policy Considerations/Recommendations} & \textbf{Computing Environment} \\
    \hline
    
    Memory-to-Memory Data Streaming Framework & 
    \begin{compactitem}
        \item Token-based authentication with time-limited credentials
        \item Separation of control and data planes
        \item Science DMZ architectures with dedicated network interfaces
        \item Templates for DSN configurations with limited user parameters 
    \end{compactitem} & 
    Development (ACE) $\rightarrow$ Production (Odo/Seeker) \\
    \hline

    Secure Scientific Service Mesh (S3M) & 
    \begin{compactitem}
        \item Centralized authentication services
        \item Fine-grained access control through standardized APIs
        \item Policy-as-code capabilities for secure access patterns
        \item Rate limiting to prevent resource exhaustion
        \item Comprehensive logging for security auditing
    \end{compactitem} & 
    Production (Odo/Seeker) $\rightarrow$ Leadership (Frontier) \\
    \hline
    
    Time-Sensitive Workload Prioritization & 
    \begin{compactitem}
        \item QoS-based preemptive scheduling tiers
        \item Resource reservation mechanisms without traditional inefficiency
        \item Project-based inheritance of QoS capabilities
        \item Formalized processes for requesting prioritized access
    \end{compactitem} & 
    All Environments \\
    \hline
    
    Multi-tenancy Resource Sharing & 
    \begin{compactitem}
        \item Containerization (Docker/OpenShift) for development
        \item Virtual machines and confidential computing enclaves for production
        \item Well-defined authorization for accessing shared resources
    \end{compactitem} & 
    Development (ACE) $\rightarrow$ Production (Odo/Seeker) \\
    \hline
    
    Cross-Facility Integration & 
    \begin{compactitem}
        \item Synchronized maintenance scheduling aligned with experimental facilities
        \item Service level agreements with defined requirements
        \item Specialized incident response for time-sensitive services
        \item Change management procedures with risk assessment
    \end{compactitem} & 
    Production (Odo/Seeker) $\rightarrow$ Leadership (Frontier) \\
    \hline

    Security Framework & 
    \begin{compactitem}
        \item Application-aware traffic governance
        \item Isolated VLANs with strict access controls
        \item Evolution from closed fortress model to selectively permeable membrane
        \item Increased testing requirements and fallback provisions
    \end{compactitem} & 
    All Environments \\
    \hline

    Resource Allocation & 
    \begin{compactitem}
        \item New evaluation criteria accounting for time sensitivity 
        \item Balance interactive workflows with broader system utilization
        \item Structured approval processes for production systems
        \item Flexible, on-demand access for development environments
    \end{compactitem} & 
    All Environments \\
    \hline

\end{tabular}
\label{tab:framework}
\end{table}

Table~\ref{tab:framework} summarizes the key technologies in our interactive HPC framework alongside their associated policy considerations. This holistic view highlights how each technological component requires specific policy adaptations across different computing environments. As workflows transition from development testbeds to production systems, these technologies and policies must evolve together, with increasing emphasis on security, reliability, and scalability. The framework provides a structured approach for researchers and system administrators to identify and address the specific requirements at each stage of the transition pathway, ensuring that interactive scientific workflows can successfully bridge the gap between experimental flexibility and production-level robustness.

\section{Conclusion}
\label{sec:conclusion}

The transition from experimental to production HPC environments represents a critical challenge for interactive scientific workflows. Through our work at OLCF and in close collaboration with many users, we have demonstrated that a carefully structured pathway, combining technological innovations with policy adaptations, can successfully bridge this gap while preserving functionality and enhancing security. Our approach addresses three core requirements: (1)~data streaming architectures that enable near real-time analysis across facility boundaries, (2)~secure authentication frameworks that balance accessibility with robust protection, and (3)~adaptive resource scheduling that accommodates time-sensitive workloads within traditional HPC environments.

The success of this transition framework depends on recognizing that each workflow has unique requirements that evolve as it progresses from development through production to leadership computing. By tailoring authentication mechanisms, API access patterns, and resource allocation strategies to each stage of this journey, researchers can maintain scientific productivity while meeting increasingly stringent security and reliability requirements. This structured progression transforms traditional batch-oriented HPC into a dynamic ecosystem capable of supporting modern scientific discovery processes that depend on interactive, time-sensitive computation.

As experimental facilities generate ever larger data volumes and scientific exploration becomes increasingly guided by machine learning models, the need for seamless integration between experimental workflows and production HPC will only grow. The transition pathway we have established provides a foundation for this integration, enabling researchers to harness the full potential of leadership computing resources for interactive scientific discovery.

\begin{credits}
\subsubsection{\ackname}
This research used resources of the Oak Ridge Leadership Computing Facility at the Oak Ridge National Laboratory, supported by the Office of Science of the U.S. Department of Energy under Contract No. DE-AC05-00OR22725.
\end{credits}

\bibliographystyle{splncs04-etal}

\end{document}